\documentclass[prb,twocolumn,a4paper]{revtex4}
\usepackage{graphicx}
\usepackage{wrapfig}
\usepackage{epstopdf}
\usepackage{float}
\usepackage{bm}
\usepackage[latin1]{inputenc}
\usepackage{placeins}
\begin{document}
\title{Transport spectroscopy of disordered graphene quantum dots \\etched into a single graphene flake}

\author{Dominikus Kölbl}
\author{Dominik M. Zumbühl}
\email{dominik.zumbuhl@unibas.ch} \affiliation{Department of Physics, University of Basel, Klingelbergstrasse 82,
CH-4056 Basel, Switzerland}
\date{\today}

\begin{abstract}    
We present transport measurements on quantum dots of sizes $45$, $60$ and $80\,$nm etched with an
$\mathrm{Ar/O_2}$-plasma into a single graphene sheet, allowing a size comparison avoiding effects from different
graphene flakes. The transport gaps and addition energies increase with decreasing dot size, as expected, and display a
strong correlation, suggesting the same physical origin for both, i.e. disorder-induced localization in presence of a
small confinement gap. Gate capacitance measurements indicate that the dot charges are located in the narrow device
region as intended. A dominant role of disorder is further substantiated by the gate dependence and the magnetic field
behavior, allowing only approximate identification of the electron-hole crossover and spin filling sequences. Finally,
we extract a g-factor consistent with $g=2$ within the error bars.
\end{abstract}
\maketitle

Spins in condensed matter systems have become an important field of research motivated by spintronics and quantum
information and the underlying fundamental physics. Graphene has several exceptional properties \cite{CastroNeto2009}
and is an exciting material promising long spin relaxation and coherence times as a result of weak spin-orbit
interaction and weak hyperfine effects due to the predominant natural abundance of the nuclear-spin free $^{12}$C
\cite{Fischer2009,Trauzettel2007}. Recent progress taking micron-scale 2D systems \cite{Novoselov2005, Zhang2005} to
nano-scale ribbons and quantum dots has opened the door to study the physics of confined charges and spins in graphene
\cite{Han2007,Todd2009,Ponomarenko2008,Liu2009,Stampfer2008a,Guettinger2010}, paving the way towards nano-device
applications. Challenges include overcoming the gapless nature of graphene \cite{Han2007,Todd2009,Ponomarenko2008},
defining tunnel barriers \cite{Liu2009}, and achieving controlled tunability of devices \cite{Stampfer2008}.

Despite these significant advances, most experiments in graphene nano-devices are currently dominated by disorder,
often masking the intrinsic (graphene) physics. Disorder is thought to arise from surface, substrate and edge
imperfections as well as intrinsic graphene defects. Investigating and suppressing disorder is therefore crucial for
further progress. Further, when studying graphene nano-devices, it is important to change the relevant parameters such
as dot size or ribbon width without significantly or qualitatively changing disorder. Here, we report electronic
transport spectroscopy of quantum dots of three different sizes fabricated on the same graphene sheet with essentially
identical disorder broadening of the Landau levels across the entire graphene flake.

\begin{figure}[tpb]
\includegraphics[width=8.7cm]{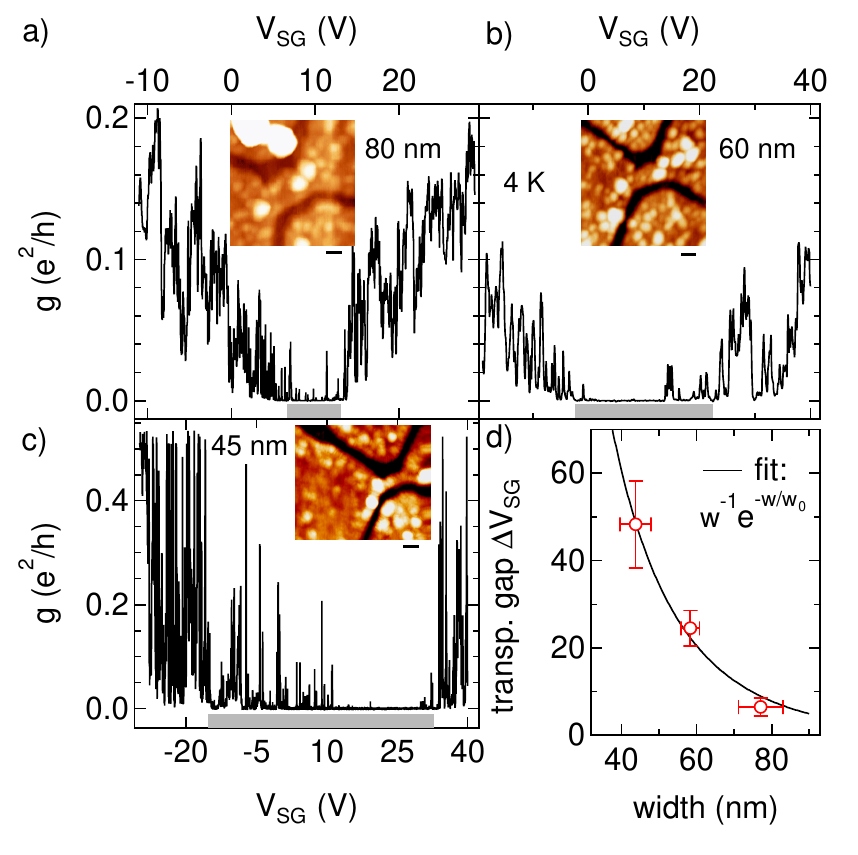}\vspace{-4mm}
\caption{\label{fig:1} Transport gaps at 4K. (a)-(c) Zero dc-bias differential conductance $g$ as a function of side
gate voltage V$_{SG}$ for each dot as labeled. The grey bars indicate the transport gap $\Delta V_{SG}$, defined as the
$V_{SG}$ range where the valley conductances remain smaller than $10^{-4}$\,e$^2$/h. The insets show AFM-images (all
scale bars 50\,nm). White speckles presumably are PMMA or other residues. (d) Transport gap $\Delta V_{SG}$ determined
from (a)-(c) (open circles) as a function of dot size. The solid curve is a fit to Ref. \cite{Sols2007}, see text. }\vspace{-4mm}
\end{figure}

The devices are approximately square-shaped graph\-ene quantum dots with designed widths $w=\,$45, 60, and 80 nm placed
on the same graphene sheet exfoliated from HOPG onto a Si wafer \cite{Novoselov2004} with a back-gate separated by a
294 nm thick oxide \cite{Blake2007}. Ti/Au (5/40\,nm) contacts are defined via standard e-beam lithography (EBL). Dots
with slightly narrower junctions to the graphene reservoirs are etched with an Ar/O$_2$ plasma using a PMMA-mask
predefined in a second EBL step. The insets in Figure 1a-c show AFM images of each dot. 
Graphene regions separated from the dots by $\sim 20\,$nm wide etched trenches are also contacted and used for side
gating the dots individually with side-gate voltage $V_{SG}$. The overall charge density can be tuned with back-gate
voltage $V_{BG}$.

Two-terminal measurements in the quantum Hall regime \cite{Williams2009} using bulk regions of the flake showed it to
be single-layer graphene. We extract a field-effect mobility of about 3'000\,cm$^2$/Vs at a density of 2\,x\,10$^{11}\,
$cm$^{-2}$ before removal of PMMA. This mobility is a lower bound as the PMMA was removed prior to the measurements
presented below. High-field Landau level broadening in the four graphene regions surrounding the three quantum dots
distributed across the $\sim 20\,\mathrm{\mu m}$ long graphene sheet was essentially identical in each region,
indicating homogeneous disorder across the entire graphene flake and therefore for all three dots, allowing a
size-comparison of transport properties without significantly changing disorder. The back-gate voltage V$_{BG} = 0$
except where stated otherwise. Measurements are performed in a dilution refrigerator unit at $T=4$\,K and $T\sim
100$\,mK (electron temperature). Conductance across the graphene nano ribbons (GNRs) is probed with a standard lock-in
technique using a small ac modulation on top of a variable dc-bias.

First, we investigate the transport gap as a function of side-gate voltage $V_{SG}$ for each dot, shown in Figure\,1a-c
at 4K. Around the charge neutrality point (CNP) located within a few volts from zero gate voltage similar for all dots,
we find a strongly suppressed conductance with sharp characteristic Coulomb blockade (CB) peaks over a wide range of
gate voltages and strong conductance fluctuations at elevated densities, both typical for GNR devices measured at low
temperatures \cite{Han2007,Ponomarenko2008,Stampfer2008a,Liu2009}. We introduce $\Delta V_{SG}$ as the $V_{SG}$-range
where the CB valley conductances remain smaller than $10^{-4}$\,e$^2$/h, as indicated by the grey bars in Figure 1a-c.
The resulting transport gap, shown in Figure\,\ref{fig:1}d, is strongly size dependent, giving larger gaps for the
smaller devices, as expected \cite{Han2007,Molitor2009,Molitor2010}.

Several theories predict the formation of a confinement gap $E_{g}$ in graphene, including tight-binding
\cite{Nakada1996}, \textsl{ab-initio} \cite{Son2006}, Anderson localization \cite{Evaldsson2008}, and many-body theory
\cite{Sols2007}, all giving similar results. The latter suggests a width $w$ dependence given by $E_{g}\sim
w^{-1}e^{-(w/w_0)}$ (with decay length $w_0$), which is widely used to analyze experimental results and also fits our
data $\Delta V_{SG}(w)$ quite well using $w_0=29.4\,\pm4.2\,nm$ (see Fig.1(d)). However, converting $\Delta V_{SG}$ to
energy ($\delta E = \alpha_{SG}\cdot\Delta V_{SG}$) using an average lever arm $\alpha_{SG} = 0.117\,\pm0.049$ eV/V
extracted from CB diamonds (see below, Fig.2(c)) results in an absolute energy scale of several $eV$, far exceeding
predictions for a simple confinement induced band gap by about two orders of magnitude. Therefore, the transport gap
$\Delta V_{SG}$ most likely is not due to geometric confinement only. Further, the appearance of numerous CB peaks
(rather than a large region of very low conductance) and conductance fluctuations surrounding the transport gap
indicate the strong influence of disorder. A large transport gap could then result from disorder localization and
Coulomb blockade in presence of a much smaller confinement gap necessary to inhibit Klein tunneling
\cite{Katsnelson2006}.

Possible sources of this disorder include graphene defects and edge disorder, trapped charges nearby, partially due to
adsorbates and PMMA residues which are clearly visible in AFM images throughout the devices (see Figure\,1, insets), as
well as other substrate and surface disorder. However, since all dots are fabricated on the same graphene sheet showing
nearly identical Landau level broadening in all regions across its length, we expect this disorder to be of similar
quality for the three dots. We note that the importance of a fabrication induced edge roughness of the order of a few
nm should increase from the 80\,nm to the 45\,nm device, where it is reaching 10\% of the device width
\cite{Evaldsson2008,Mucciolo2009}.
\begin{figure}[tpb]
\includegraphics[width=8.7cm]{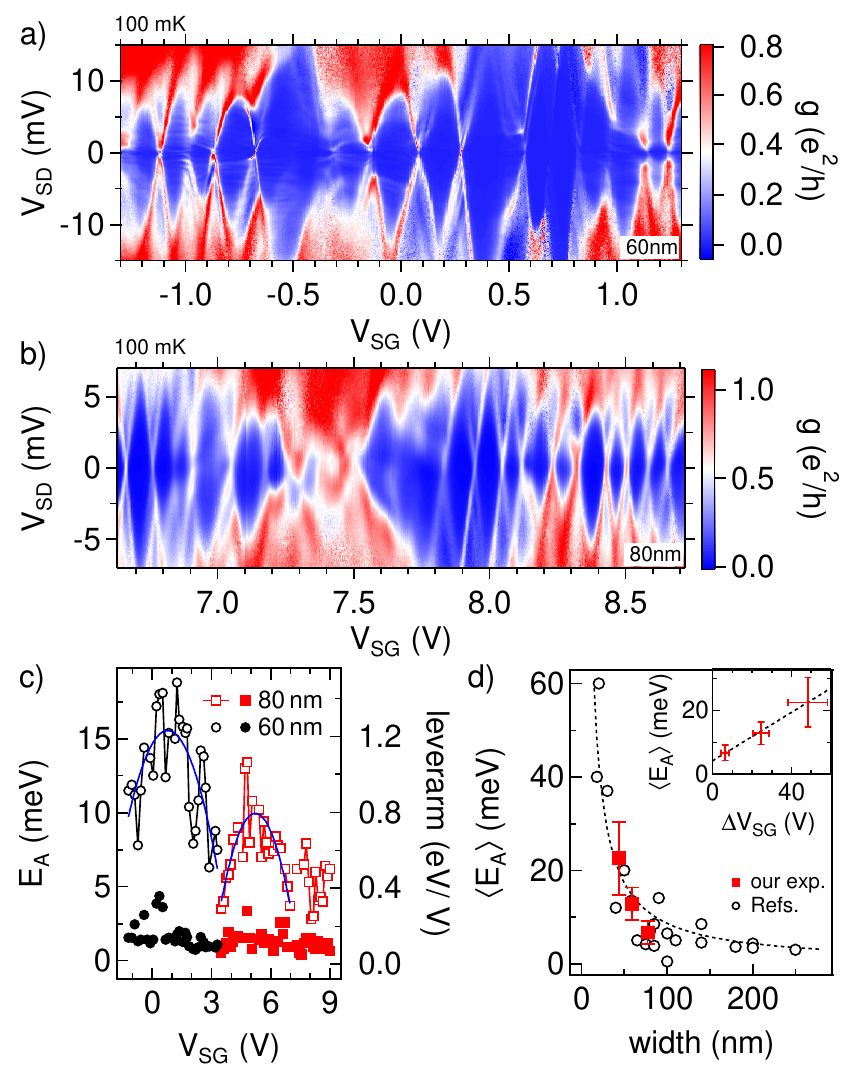}\vspace{-4mm}
\caption{\label{fig:2} Coulomb diamonds (a),(b) Differential conductance (color scale) as a function of source-drain
voltage $V_{SD}$ and side-gate voltage $V_{SG}$ at $T \sim $ 100 mK of the 60 and 80\,nm dots, as labeled. (c) Addition
energies (left axis, open symbols) and corresponding $V_{SG}$ lever arms (right axis, filled symbols) from Coulomb
diamonds as in (a) and (b) but with extended $V_{SG}$ range. Blue curves (parabolas) are shown as a guide to the eye,
indicating a peak in $E_{A}$ at the CNP for both dots, also confirmed by $B_\bot$ data, see text. (d) Size-dependence
of the average addition energy $\langle E_{A}\rangle$ (squares). Error bars denote standard deviation. Circles are from
Refs. \cite{Ponomarenko2008,Stampfer2008,Schnez2009,Stampfer2009,Molitor2009,Guettinger2009,Liu2009,
Neubeck2010,Wang2011a}. The dashed curve is an $\epsilon/w$ fit with $\epsilon = 870\pm70\,$meV\,nm, see text. The
inset shows $\langle E_{A}\rangle$ vs. $\Delta V_{SG}$ with a line fit (dashed black) indicating a strong correlation.}\vspace{-4mm}
\end{figure}

Clear Coulomb diamonds are seen in finite bias measurements for all three dots, shown in Figure\,2a,b for the larger
dots at 100\,mK, indicating the formation of a tunnel coupled quantum dot in the transport gap region. We find
signatures of excited states in sequential tunneling (typically at $\sim$meV energies), but also cotunneling features.
We extract the addition energies $E_A$ and $V_{SG}$ lever arms from similar data extending over a larger $V_{SG}$-range
for the two larger dots, shown in Figure\,2c. Both dots show similar lever arms, as expected due to similar geometry,
roughly independent of $V_{SG}$. The addition energies are larger in the smaller dot, on average, as expected. Further,
a maximum in $E_{A}$ as a function of $V_{SG}$ -- indicated by the blue curves -- is seen close to the bulk CNP,
roughly marking the electron-to-hole crossover. However, we cannot identify the zero-occupation diamond and the
absolute charge-number in these dots, though the expected confinement-induced band gap \cite{Son2006} is comparable to
the observed addition energies.

The size dependence of the average addition energy $\langle E_{A}\rangle$ obtained from Coulomb diamond measurements
over a large gate voltage range is shown in Figure\,2d (red squares), in good agreement with previous reports of
similar size devices (black circles)
\cite{Ponomarenko2008,Stampfer2008,Schnez2009,Stampfer2009,Liu2009,Molitor2009,Guettinger2009,Neubeck2010,Wang2011}. A
fit to the single dot theory $\langle E_{A} \rangle = \epsilon/w$ \cite{Sols2007} gives decent agreement (see dashed
curve), resulting in $\epsilon = 870\,\pm70\,$meV\,nm, comparable with other experiments \cite{Han2007,Molitor2010}.
Interestingly, for the present three dots, we find a clear correlation between the average addition energy $\langle
E_{A}\rangle$ and the transport gap size $\Delta V_{SG}$ (see inset Figure\,2d), suggesting the same physical origin
for both energy scales.

Further, we can estimate the effective dot area via the back-gate capacitance taken from diamonds and using a simple
parallel plate capacitor model \cite{PlateCap}. This simple model should give a good estimate of the area for the
larger devices, where the etched trenches defining the dots in the otherwise continuous graphene layer are narrow
compared to the device diameter. We note that the simple plate capacitor model used here for the back-gate does not
apply to the total capacitance, which is significantly larger than the back-gate capacitance. The extracted areas agree
well (within the error bars of $\sim15\,\%$) with the actual dot sizes (from AFM scans) for the two larger devices,
suggesting that the electrons are indeed located in the lithographically intended region of the GNR. Overall, the above
results seem to indicate predominant formation of single quantum dots in these devices.

However, we also find a number of overlapping diamonds or diamonds that do not close at low bias, indicating formation
of double or multiple dots \cite{Droescher2011} in a repeatable way (during the same cool down) as a function of gate
voltage. This is further substantiated by $V_{SG}$ and $V_{BG}$ scans shown in Figure\,3a,b. Regions in gate space of
parallel lines with a fixed slope (given by the relative side- and back-gate leverarms) characteristic for a single dot
are alternating with non-parallel, honey-comb like features \cite{Todd2009,Stampfer2009,Molitor2009a} (again repeatable
in gate voltage), indicating double or multiple-dot formation \cite{DDreview}, presumably as a result of the pronounced
disorder potential. As gate voltage is changed monotonously, the dot appears to sporadically rearrange its geometry,
deforming between a simple, single dot and more complicated configurations.

\begin{figure}[tpb]
\includegraphics[width=8.7cm]{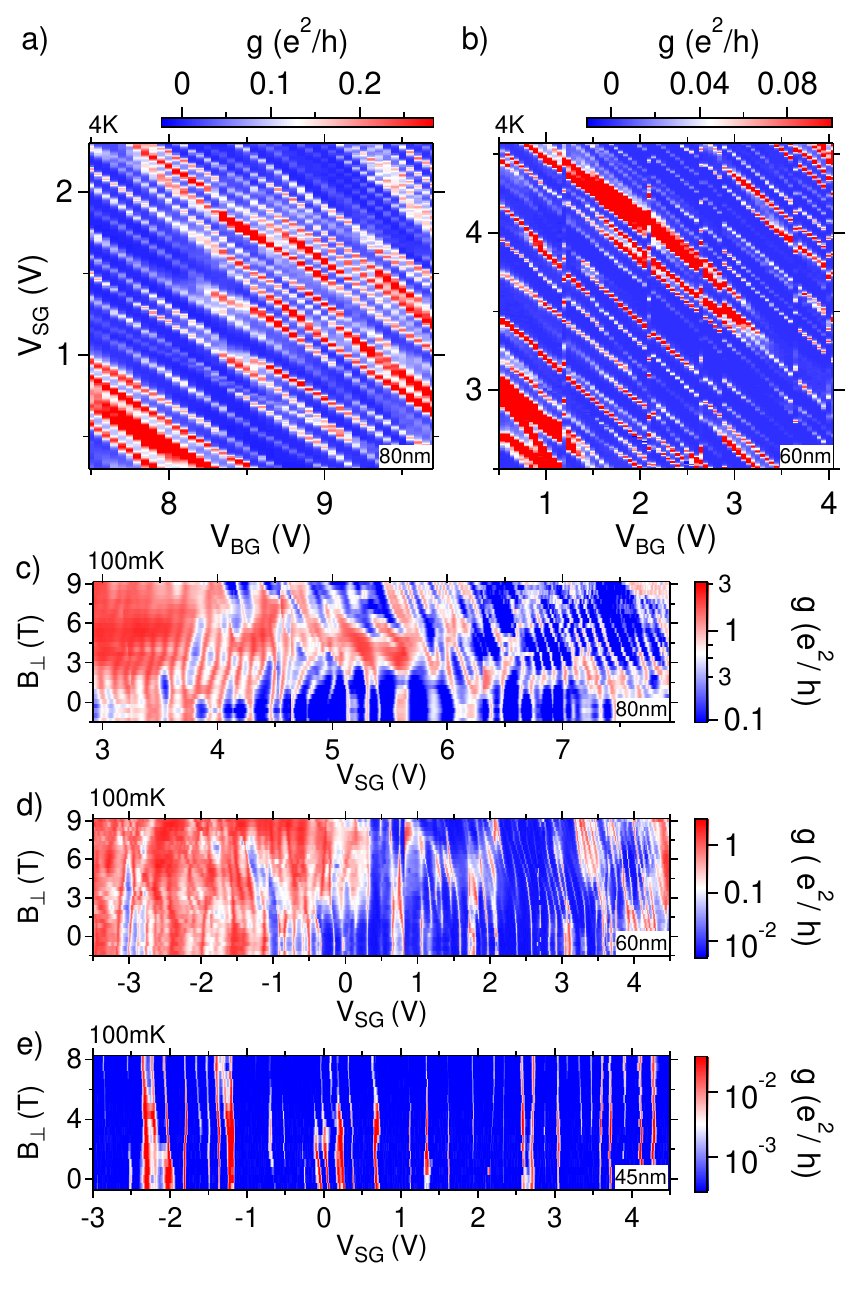}\vspace{-4mm}
\caption{\label{fig:3}Gate-gate sweeps and $B_\bot$ peak motion: (a),(b) Differential conductance showing CB peaks as a
function of $V_{SG}$ and $V_{BG}$ at 4\,K. Parallel lines indicating single dot behavior alternate with bending and
merging features that can arise from multiple-dot formation (repeatable). (c)-(e) CB peak evolution in a perpendicular
field $B_\bot$ of the 80, 60 and 45\,nm dots at 100\,mK. At large $B_\bot$, peaks bend towards the graphene zero
Landau-level around the electron-hole transition ($V_{SG}\sim 5$\,V for 80\,nm dot, and $V_{SG}\sim 1$\,V for 60\,nm
dot), more clearly visible for the larger dots, see text.}\vspace{-4mm}
\end{figure}

We now turn to perpendicular magnetic field $B_\bot$ measurements, shown in Figure\,3c-e for all three dots at
$V_{SD}=0$ and $T=100$\,mK. Besides a strong variation of the peak conductance, the peak positions of the 80\,nm-device
bend towards $V_{SG} \sim 5\,$V for large $B_\bot$, as expected for the $0th$ graphene Landau level at the CNP
\cite{Recher2009,Guettinger2009,Libisch2010}. Therefore, we can extract the CNP in this device to be located around
5\,V, consistent with the highest value of $E_{A}$ found for $V_{SG}$ = 4.8\,V (Figure\,2c). Similarly, for the 60\,nm
device, the electron-hole crossover is found around $V_{SG} \sim 1\,$V, again consistent with the previously determined
maximal $E_A$ at $V_{SG} = 1.25\,$V (Figure\,2c), though for this devices the $B_\bot$ bending of the peaks is weaker.
Therefore, the CNPs in both dots are separated by only a few Volts, both close to zero. Landau level bending becomes
visible at high fields when the magnetic length $l_{B} = \sqrt{\hbar/eB}$ is much smaller than the device size $w$
\cite{Recher2009,Libisch2010}, making the effect weakest in the smallest dot (Figure\,3e).

Beyond Landau levels, paired peak motion due to consecutive filling of the same orbital with opposite spins (spin
pairs) can also be observed in the $B_\bot$ dependence. Periods of four were not identifiable, suggesting a broken
valley degeneracy in these dots. Spin pairs are most clearly visible for the largest device (where the B$_{\bot}$
effect is most pronounced), where some pairs particularly at high electron/hole densities away from the CNP exhibit
reproducible parallel evolution over a significant range in $B_{\bot}$, see e.g. Figure\,3c, 3\,V $< V_{SG} <$ 4.5\,V.
However, the low-density region around the CNP which is more strongly affected by disorder \cite{Libisch2010} appears
more complicated and clear pairs could not be found, similar to the smaller dots, which are also more weakly coupled to
the reservoirs. These efforts are further hampered by disorder driven dot rearrangements (single to double dot
transitions as a function of $V_{SG}$) as described before and sporadic switching in gate voltage observed in these
devices.

\begin{figure}[tpb]
\includegraphics[width=8.3cm]{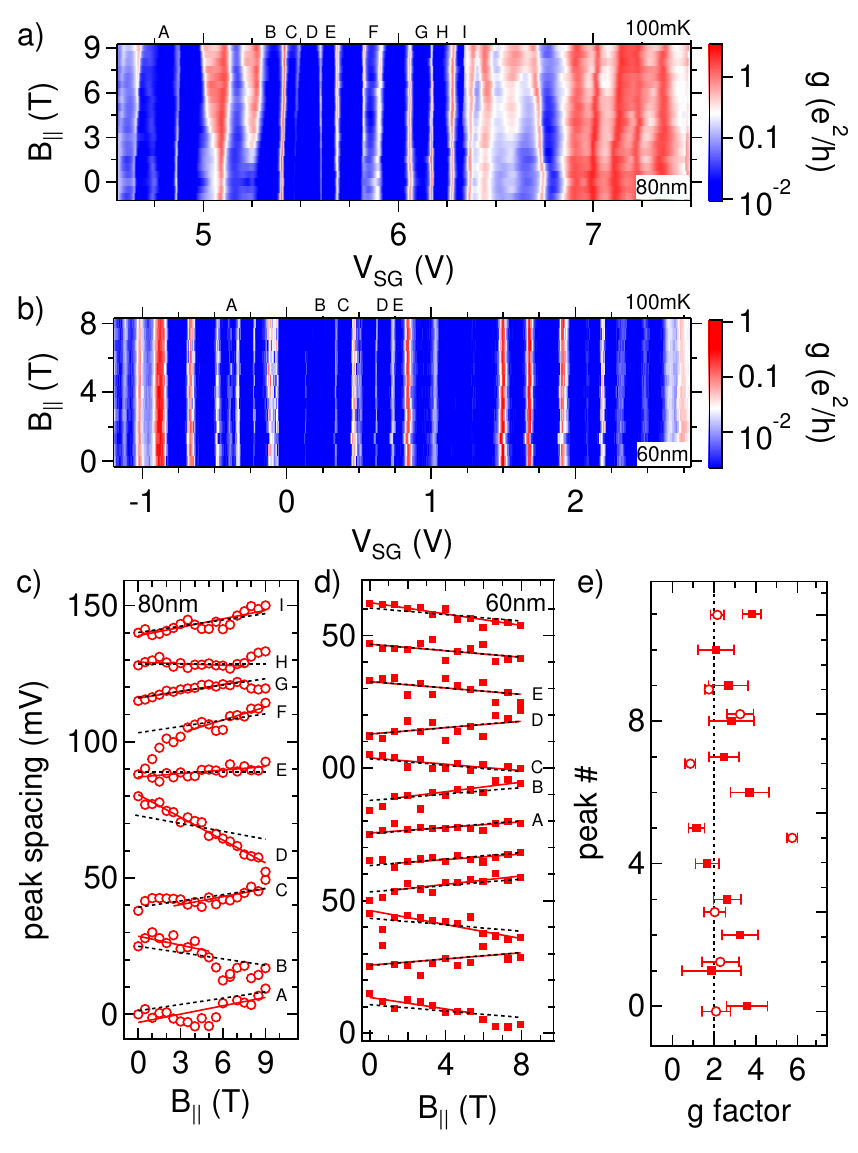}\vspace{-4mm}
\caption{\label{fig:4} $B_\parallel$ peak motion and g-factors: (a),(b) Differential conductance as a function of
$B_\parallel$ and $V_{SG}$ for the larger dots at 100\,mK. (c),(d) Selection of CB peak spacings which are
approximately linear in $B_\parallel$, offset for clarity, see text. Solid red lines are best linear fits. The extent
of each line is the fit range used. For comparison, the $\pm g \mu_B$ and zero slopes are also shown (dashed black
lines).  (e) Summary of g-factors from (c) and (d) (excluding zero-slope data), with open circles for 80\,nm dot and
solid squares for 60\,nm dot. Vertical positions are chosen to align with data in (c) and (d). Error bars are standard
deviations from linear fits. }\vspace{-4mm}
\end{figure}

The evolution of the CB peak spacing in an in-plane magnetic field $B_\parallel$ can in principle reflect the spin
filling sequence \cite{Folk2001}. In graphene quantum dots, spin-orbit coupling can be assumed to be very weak and the
Land\'{e} g-factor $g\sim2$. For dot-diameters $d<100\,$nm, the orbital level spacing $\Delta > 2\,$meV remains larger
then the Zeeman splitting $E_Z=g\mu_B B$ (with Bohr magneton $\mu_B$) for $B_\parallel \leq10\,$T. In this case, and if
electron-electron interactions are negligible, one might expect a simple alternating Pauli spin sequence giving peak
spacings which increase or decrease with slope $g\mu_B$. $B_\parallel$-independent peak spacings (slope zero), however,
would be absent in this simple picture, since these indicate a filling of two subsequent identical spins induced by
interactions \cite{Folk2001}, resulting in total spin $S>1/2$.

Figure\,4a,b shows the CB peak positions of the 80\,nm and 60\,nm devices at 100\,mK as a function of $B_\parallel$
(separate cool down) over a range of $V_{SG}$ including the electron-hole transitions. We fit Gaussians to the CB peaks
to obtain the peak positions and evaluate the peak spacing as a function of $B_\parallel$. While some peak spacings
show the expected slopes (zero or $\pm g \mu_B$), others exhibit more complicated, nonlinear $B_\parallel$ dependence.
This could be due to disorder driven dot-rearrangements as a function of $V_{SG}$ as mentioned above, a slight
$B_\parallel$ misalignment with a resulting $B_\bot$ component of $B_\parallel$ (a few degrees here) or other orbital
coupling of $B_\parallel$, e.g. by threading flux through the graphene surface ripples \cite{Lundeberg2010}.

In an attempt to avoid these $B_\parallel$ complications, we select peaks spacings approximately linear over a
sufficiently large range of $B_\parallel$ without rejecting any slope, plotted in Figure\,4c,d (offset for clarity),
also labeled A-I and $\alpha$-$\epsilon$ above their corresponding peaks in Figure\,4a,b. We extract the slopes with
best fits (solid red lines) and also indicate the closest standard slope (dashed black lines, slopes $0$, $\pm g
\mu_B$) for comparison, using the average lever arms previously measured from Coulomb diamonds of each dot. The
resulting g-factors are summarized in Figure\,4e for both dots. While in several cases, good agreement with the
expected $g\sim2$ is found (see e.g. A,B,C,G,I), we also notice horizontal, $B_\parallel$ independent peak spacings
indicating non-trivial spin filling (e.g. H). Further, slopes strongly deviating from $g=2$ are also seen, which is not
surprising considering the $B_\parallel$ issues mentioned before. Nevertheless, averaging over the data in Figure\,4e
from both dots, we obtain $g=2.7\pm1.1$ (excluding the obvious near-zero point H), consistent with $g=2$ as expected
and in line with other experiments \cite{Lundeberg2009,Guettinger2010,Rao2011}.

In summary, we have presented transport spectroscopy of graphene quantum dots on the same graphene flake with nearly
identical disorder broadening. This allows a size comparison without changing disorder, displaying the expected size
dependence of transport gap and addition energy as well as clear correlation between both, suggesting disorder induced
localization in presence of a confinement gap as the physical origin for both effects. Gate capacitance measurements
indicate that the dot charges are located in the narrow device region as expected. Even though the electron-hole
transitions could not be precisely located ($\pm$ few electrons) and the spin filling sequences were not fully
tractable, both ultimately due to disorder, the average g-factor is consistent with $g=2$, though with significant
error bar.

Overall, the combined data clearly draw a consistent picture of pronounced disorder effects which are masking the
interesting low-density, few electron regime in these graphene devices. For future nano-graphene experiments, it will
therefore be very important to investigate and suppress disorder, e.g. by removal of substrate \cite{Tombros2011} and
adsorbate disorder, by a high degree of control over the graphene edges \cite{Moreno2009,Shi2011,Wang2011} and
elimination of any residual intrinsic graphene defects.

The authors thank G. Burkhard and B. Trauzettel for helpful discussions. This work was supported by the Swiss
Nanoscience Institute (SNI), Swiss NSF, NCCR QSIT and an ERC starting grant.


\begin{thebibliography}{42}
\expandafter\ifx\csname natexlab\endcsname\relax\def\natexlab#1{#1}\fi
\expandafter\ifx\csname bibnamefont\endcsname\relax
  \def\bibnamefont#1{#1}\fi
\expandafter\ifx\csname bibfnamefont\endcsname\relax
  \def\bibfnamefont#1{#1}\fi
\expandafter\ifx\csname citenamefont\endcsname\relax
  \def\citenamefont#1{#1}\fi
\expandafter\ifx\csname url\endcsname\relax
  \def\url#1{\texttt{#1}}\fi
\expandafter\ifx\csname urlprefix\endcsname\relax\def\urlprefix{URL }\fi
\providecommand{\bibinfo}[2]{#2}
\providecommand{\eprint}[2][]{\url{#2}}

\bibitem[{\citenamefont{Castro~Neto et~al.}(2009)\citenamefont{Castro~Neto,
  Guinea, Peres, Novoselov, and Geim}}]{CastroNeto2009}
\bibinfo{author}{\bibfnamefont{A.~H.} \bibnamefont{Castro~Neto}},
  \bibinfo{author}{\bibfnamefont{F.}~\bibnamefont{Guinea}},
  \bibinfo{author}{\bibfnamefont{N.~M.~R.} \bibnamefont{Peres}},
  \bibinfo{author}{\bibfnamefont{K.~S.} \bibnamefont{Novoselov}},
  \bibnamefont{and} \bibinfo{author}{\bibfnamefont{A.~K.} \bibnamefont{Geim}},
  \bibinfo{journal}{Rev. Mod. Phys.} \textbf{\bibinfo{volume}{81}},
  \bibinfo{pages}{109} (\bibinfo{year}{2009}).

\bibitem[{\citenamefont{Fischer et~al.}(2009)\citenamefont{Fischer, Trauzettel,
  and Loss}}]{Fischer2009}
\bibinfo{author}{\bibfnamefont{J.}~\bibnamefont{Fischer}},
  \bibinfo{author}{\bibfnamefont{B.}~\bibnamefont{Trauzettel}},
  \bibnamefont{and} \bibinfo{author}{\bibfnamefont{D.}~\bibnamefont{Loss}},
  \bibinfo{journal}{Phys. Rev. B} \textbf{\bibinfo{volume}{80}},
  \bibinfo{pages}{155401} (\bibinfo{year}{2009}).

\bibitem[{\citenamefont{Trauzettel et~al.}(2007)\citenamefont{Trauzettel,
  Bulaev, Loss, and Burkard}}]{Trauzettel2007}
\bibinfo{author}{\bibfnamefont{B.}~\bibnamefont{Trauzettel}},
  \bibinfo{author}{\bibfnamefont{D.~V.} \bibnamefont{Bulaev}},
  \bibinfo{author}{\bibfnamefont{D.}~\bibnamefont{Loss}}, \bibnamefont{and}
  \bibinfo{author}{\bibfnamefont{G.}~\bibnamefont{Burkard}},
  \bibinfo{journal}{Nat. Phys.} \textbf{\bibinfo{volume}{3}},
  \bibinfo{pages}{192} (\bibinfo{year}{2007}).

\bibitem[{\citenamefont{Novoselov et~al.}(2005)\citenamefont{Novoselov, Geim,
  Morozov, Jiang, Katsnelson, Grigorieva, Dubonos, and Firsov}}]{Novoselov2005}
\bibinfo{author}{\bibfnamefont{K.~S.} \bibnamefont{Novoselov}},
  \bibinfo{author}{\bibfnamefont{A.~K.} \bibnamefont{Geim}},
  \bibinfo{author}{\bibfnamefont{S.~V.} \bibnamefont{Morozov}},
  \bibinfo{author}{\bibfnamefont{D.}~\bibnamefont{Jiang}},
  \bibinfo{author}{\bibfnamefont{M.~I.} \bibnamefont{Katsnelson}},
  \bibinfo{author}{\bibfnamefont{I.~V.} \bibnamefont{Grigorieva}},
  \bibinfo{author}{\bibfnamefont{S.~V.} \bibnamefont{Dubonos}},
  \bibnamefont{and} \bibinfo{author}{\bibfnamefont{A.~A.}
  \bibnamefont{Firsov}}, \bibinfo{journal}{Nature}
  \textbf{\bibinfo{volume}{438}}, \bibinfo{pages}{197} (\bibinfo{year}{2005}).

\bibitem[{\citenamefont{Zhang et~al.}(2005)\citenamefont{Zhang, Tan, Stormer,
  and Kim}}]{Zhang2005}
\bibinfo{author}{\bibfnamefont{Y.}~\bibnamefont{Zhang}},
  \bibinfo{author}{\bibfnamefont{Y.-W.} \bibnamefont{Tan}},
  \bibinfo{author}{\bibfnamefont{H.~L.} \bibnamefont{Stormer}},
  \bibnamefont{and} \bibinfo{author}{\bibfnamefont{P.}~\bibnamefont{Kim}},
  \bibinfo{journal}{Nature} \textbf{\bibinfo{volume}{438}},
  \bibinfo{pages}{201} (\bibinfo{year}{2005}).

\bibitem[{\citenamefont{Han et~al.}(2007)\citenamefont{Han, Özyilmaz, Zhang,
  and Kim}}]{Han2007}
\bibinfo{author}{\bibfnamefont{M.~Y.} \bibnamefont{Han}},
  \bibinfo{author}{\bibfnamefont{B.}~\bibnamefont{Özyilmaz}},
  \bibinfo{author}{\bibfnamefont{Y.}~\bibnamefont{Zhang}}, \bibnamefont{and}
  \bibinfo{author}{\bibfnamefont{P.}~\bibnamefont{Kim}},
  \bibinfo{journal}{Phys. Rev. Lett.} \textbf{\bibinfo{volume}{98}},
  \bibinfo{pages}{206805} (\bibinfo{year}{2007}).

\bibitem[{\citenamefont{Todd et~al.}(2009)\citenamefont{Todd, Chou, Amasha, and
  Goldhaber-Gordon}}]{Todd2009}
\bibinfo{author}{\bibfnamefont{K.}~\bibnamefont{Todd}},
  \bibinfo{author}{\bibfnamefont{H.}~\bibnamefont{Chou}},
  \bibinfo{author}{\bibfnamefont{S.}~\bibnamefont{Amasha}}, \bibnamefont{and}
  \bibinfo{author}{\bibfnamefont{D.}~\bibnamefont{Goldhaber-Gordon}},
  \bibinfo{journal}{Nano Lett.} \textbf{\bibinfo{volume}{9}},
  \bibinfo{pages}{416} (\bibinfo{year}{2009}).

\bibitem[{\citenamefont{Ponomarenko et~al.}(2008)\citenamefont{Ponomarenko,
  Schedin, Katsnelson, Yang, Hill, Novoselov, and Geim}}]{Ponomarenko2008}
\bibinfo{author}{\bibfnamefont{L.~A.} \bibnamefont{Ponomarenko}},
  \bibinfo{author}{\bibfnamefont{F.}~\bibnamefont{Schedin}},
  \bibinfo{author}{\bibfnamefont{M.~I.} \bibnamefont{Katsnelson}},
  \bibinfo{author}{\bibfnamefont{R.}~\bibnamefont{Yang}},
  \bibinfo{author}{\bibfnamefont{E.~W.} \bibnamefont{Hill}},
  \bibinfo{author}{\bibfnamefont{K.~S.} \bibnamefont{Novoselov}},
  \bibnamefont{and} \bibinfo{author}{\bibfnamefont{A.~K.} \bibnamefont{Geim}},
  \bibinfo{journal}{Science} \textbf{\bibinfo{volume}{320}},
  \bibinfo{pages}{356} (\bibinfo{year}{2008}).

\bibitem[{\citenamefont{Liu et~al.}(2009)\citenamefont{Liu, Oostinga, Morpurgo,
  and Vandersypen}}]{Liu2009}
\bibinfo{author}{\bibfnamefont{X.}~\bibnamefont{Liu}},
  \bibinfo{author}{\bibfnamefont{J.~B.} \bibnamefont{Oostinga}},
  \bibinfo{author}{\bibfnamefont{A.~F.} \bibnamefont{Morpurgo}},
  \bibnamefont{and} \bibinfo{author}{\bibfnamefont{L.~M.~K.}
  \bibnamefont{Vandersypen}}, \bibinfo{journal}{Phys. Rev. B.}
  \textbf{\bibinfo{volume}{80}}, \bibinfo{pages}{121407}
  (\bibinfo{year}{2009}).

\bibitem[{\citenamefont{Stampfer
  et~al.}(2008{\natexlab{a}})\citenamefont{Stampfer, Schurtenberger, Molitor,
  Güttinger, Ihn, and Ensslin}}]{Stampfer2008a}
\bibinfo{author}{\bibfnamefont{C.}~\bibnamefont{Stampfer}},
  \bibinfo{author}{\bibfnamefont{E.}~\bibnamefont{Schurtenberger}},
  \bibinfo{author}{\bibfnamefont{F.}~\bibnamefont{Molitor}},
  \bibinfo{author}{\bibfnamefont{J.}~\bibnamefont{Güttinger}},
  \bibinfo{author}{\bibfnamefont{T.}~\bibnamefont{Ihn}}, \bibnamefont{and}
  \bibinfo{author}{\bibfnamefont{K.}~\bibnamefont{Ensslin}},
  \bibinfo{journal}{Nano Lett.} \textbf{\bibinfo{volume}{8}},
  \bibinfo{pages}{2378} (\bibinfo{year}{2008}{\natexlab{a}}).

\bibitem[{\citenamefont{Güttinger et~al.}(2010)\citenamefont{Güttinger, Frey,
  Stampfer, Ihn, and Ensslin}}]{Guettinger2010}
\bibinfo{author}{\bibfnamefont{J.}~\bibnamefont{Güttinger}},
  \bibinfo{author}{\bibfnamefont{T.}~\bibnamefont{Frey}},
  \bibinfo{author}{\bibfnamefont{C.}~\bibnamefont{Stampfer}},
  \bibinfo{author}{\bibfnamefont{T.}~\bibnamefont{Ihn}}, \bibnamefont{and}
  \bibinfo{author}{\bibfnamefont{K.}~\bibnamefont{Ensslin}},
  \bibinfo{journal}{Phys. Rev. Lett.} \textbf{\bibinfo{volume}{105}},
  \bibinfo{pages}{116801} (\bibinfo{year}{2010}).

\bibitem[{\citenamefont{Stampfer
  et~al.}(2008{\natexlab{b}})\citenamefont{Stampfer, Güttinger, Molitor, Graf,
  Ihn, and Ensslin}}]{Stampfer2008}
\bibinfo{author}{\bibfnamefont{C.}~\bibnamefont{Stampfer}},
  \bibinfo{author}{\bibfnamefont{J.}~\bibnamefont{Güttinger}},
  \bibinfo{author}{\bibfnamefont{F.}~\bibnamefont{Molitor}},
  \bibinfo{author}{\bibfnamefont{D.}~\bibnamefont{Graf}},
  \bibinfo{author}{\bibfnamefont{T.}~\bibnamefont{Ihn}}, \bibnamefont{and}
  \bibinfo{author}{\bibfnamefont{K.}~\bibnamefont{Ensslin}},
  \bibinfo{journal}{Appl. Phys. Lett.} \textbf{\bibinfo{volume}{92}},
  \bibinfo{pages}{012102} (\bibinfo{year}{2008}{\natexlab{b}}).

\bibitem[{\citenamefont{Sols et~al.}(2007)\citenamefont{Sols, Guinea, and
  Castro~Neto}}]{Sols2007}
\bibinfo{author}{\bibfnamefont{F.}~\bibnamefont{Sols}},
  \bibinfo{author}{\bibfnamefont{F.}~\bibnamefont{Guinea}}, \bibnamefont{and}
  \bibinfo{author}{\bibfnamefont{A.~H.} \bibnamefont{Castro~Neto}},
  \bibinfo{journal}{Phys. Rev. Lett.} \textbf{\bibinfo{volume}{99}},
  \bibinfo{pages}{166803} (\bibinfo{year}{2007}).

\bibitem[{\citenamefont{Novoselov et~al.}(2004)\citenamefont{Novoselov, Geim,
  Morozov, Jiang, Zhang, Dubonos, Grigorieva, and Firsov}}]{Novoselov2004}
\bibinfo{author}{\bibfnamefont{K.~S.} \bibnamefont{Novoselov}},
  \bibinfo{author}{\bibfnamefont{A.}~\bibnamefont{Geim}},
  \bibinfo{author}{\bibfnamefont{S.~V.} \bibnamefont{Morozov}},
  \bibinfo{author}{\bibfnamefont{D.}~\bibnamefont{Jiang}},
  \bibinfo{author}{\bibfnamefont{Y.}~\bibnamefont{Zhang}},
  \bibinfo{author}{\bibfnamefont{S.}~\bibnamefont{Dubonos}},
  \bibinfo{author}{\bibfnamefont{I.}~\bibnamefont{Grigorieva}},
  \bibnamefont{and} \bibinfo{author}{\bibfnamefont{A.~A.}
  \bibnamefont{Firsov}}, \bibinfo{journal}{Science}
  \textbf{\bibinfo{volume}{306}}, \bibinfo{pages}{5696} (\bibinfo{year}{2004}).

\bibitem[{\citenamefont{Blake et~al.}(2007)\citenamefont{Blake, Hill,
  Castro~Neto, Novoselov, Jiang, Yang, Booth, and Geim}}]{Blake2007}
\bibinfo{author}{\bibfnamefont{P.}~\bibnamefont{Blake}},
  \bibinfo{author}{\bibfnamefont{E.~W.} \bibnamefont{Hill}},
  \bibinfo{author}{\bibfnamefont{A.~H.} \bibnamefont{Castro~Neto}},
  \bibinfo{author}{\bibfnamefont{K.~S.} \bibnamefont{Novoselov}},
  \bibinfo{author}{\bibfnamefont{D.}~\bibnamefont{Jiang}},
  \bibinfo{author}{\bibfnamefont{R.}~\bibnamefont{Yang}},
  \bibinfo{author}{\bibfnamefont{T.~J.} \bibnamefont{Booth}}, \bibnamefont{and}
  \bibinfo{author}{\bibfnamefont{A.~K.} \bibnamefont{Geim}},
  \bibinfo{journal}{Appl. Phys. Lett.} \textbf{\bibinfo{volume}{91}},
  \bibinfo{pages}{063124} (\bibinfo{year}{2007}).

\bibitem[{\citenamefont{Williams et~al.}(2009)\citenamefont{Williams, Abanin,
  Dicarlo, Levitov, and Marcus}}]{Williams2009}
\bibinfo{author}{\bibfnamefont{J.~R.} \bibnamefont{Williams}},
  \bibinfo{author}{\bibfnamefont{D.~A.} \bibnamefont{Abanin}},
  \bibinfo{author}{\bibfnamefont{L.}~\bibnamefont{Dicarlo}},
  \bibinfo{author}{\bibfnamefont{L.~S.} \bibnamefont{Levitov}},
  \bibnamefont{and} \bibinfo{author}{\bibfnamefont{C.~M.}
  \bibnamefont{Marcus}}, \bibinfo{journal}{Phys. Rev. B.}
  \textbf{\bibinfo{volume}{80}}, \bibinfo{pages}{045408}
  (\bibinfo{year}{2009}).

\bibitem[{\citenamefont{Molitor
  et~al.}(2009{\natexlab{a}})\citenamefont{Molitor, Jacobsen, Stampfer,
  Güttinger, Ihn, and Ensslin}}]{Molitor2009}
\bibinfo{author}{\bibfnamefont{F.}~\bibnamefont{Molitor}},
  \bibinfo{author}{\bibfnamefont{A.}~\bibnamefont{Jacobsen}},
  \bibinfo{author}{\bibfnamefont{C.}~\bibnamefont{Stampfer}},
  \bibinfo{author}{\bibfnamefont{J.}~\bibnamefont{Güttinger}},
  \bibinfo{author}{\bibfnamefont{T.}~\bibnamefont{Ihn}}, \bibnamefont{and}
  \bibinfo{author}{\bibfnamefont{K.}~\bibnamefont{Ensslin}},
  \bibinfo{journal}{Phys. Rev. B.} \textbf{\bibinfo{volume}{79}},
  \bibinfo{pages}{075426} (\bibinfo{year}{2009}{\natexlab{a}}).

\bibitem[{\citenamefont{Molitor et~al.}(2010)\citenamefont{Molitor, Stampfer,
  Güttinger, Jacobsen, Ihn, and Ensslin}}]{Molitor2010}
\bibinfo{author}{\bibfnamefont{F.}~\bibnamefont{Molitor}},
  \bibinfo{author}{\bibfnamefont{C.}~\bibnamefont{Stampfer}},
  \bibinfo{author}{\bibfnamefont{J.}~\bibnamefont{Güttinger}},
  \bibinfo{author}{\bibfnamefont{A.}~\bibnamefont{Jacobsen}},
  \bibinfo{author}{\bibfnamefont{T.}~\bibnamefont{Ihn}}, \bibnamefont{and}
  \bibinfo{author}{\bibfnamefont{K.}~\bibnamefont{Ensslin}},
  \bibinfo{journal}{Semicond. Sci. Technol.} \textbf{\bibinfo{volume}{25}},
  \bibinfo{pages}{034002} (\bibinfo{year}{2010}).

\bibitem[{\citenamefont{Nakada et~al.}(1996)\citenamefont{Nakada, Fujita,
  Dresselhaus, and Dresselhaus}}]{Nakada1996}
\bibinfo{author}{\bibfnamefont{K.}~\bibnamefont{Nakada}},
  \bibinfo{author}{\bibfnamefont{M.}~\bibnamefont{Fujita}},
  \bibinfo{author}{\bibfnamefont{G.}~\bibnamefont{Dresselhaus}},
  \bibnamefont{and} \bibinfo{author}{\bibfnamefont{M.~S.}
  \bibnamefont{Dresselhaus}}, \bibinfo{journal}{Phys. Rev. B}
  \textbf{\bibinfo{volume}{54}}, \bibinfo{pages}{17954} (\bibinfo{year}{1996}).

\bibitem[{\citenamefont{Son et~al.}(2006)\citenamefont{Son, Cohen, and
  Louie}}]{Son2006}
\bibinfo{author}{\bibfnamefont{Y.~W.} \bibnamefont{Son}},
  \bibinfo{author}{\bibfnamefont{M.~L.} \bibnamefont{Cohen}}, \bibnamefont{and}
  \bibinfo{author}{\bibfnamefont{S.~G.} \bibnamefont{Louie}},
  \bibinfo{journal}{Phys. Rev. Lett.} \textbf{\bibinfo{volume}{97}},
  \bibinfo{pages}{216803} (\bibinfo{year}{2006}).

\bibitem[{\citenamefont{Evaldsson et~al.}(2008)\citenamefont{Evaldsson,
  Zozoulenko, Xu, and Heinzel}}]{Evaldsson2008}
\bibinfo{author}{\bibfnamefont{M.}~\bibnamefont{Evaldsson}},
  \bibinfo{author}{\bibfnamefont{I.}~\bibnamefont{Zozoulenko}},
  \bibinfo{author}{\bibfnamefont{H.}~\bibnamefont{Xu}}, \bibnamefont{and}
  \bibinfo{author}{\bibfnamefont{T.}~\bibnamefont{Heinzel}},
  \bibinfo{journal}{Phys. Rev. B.} \textbf{\bibinfo{volume}{78}},
  \bibinfo{pages}{161407} (\bibinfo{year}{2008}).

\bibitem[{\citenamefont{Katsnelson et~al.}(2006)\citenamefont{Katsnelson,
  Novoselov, and Geim}}]{Katsnelson2006}
\bibinfo{author}{\bibfnamefont{M.~I.} \bibnamefont{Katsnelson}},
  \bibinfo{author}{\bibfnamefont{K.~S.} \bibnamefont{Novoselov}},
  \bibnamefont{and} \bibinfo{author}{\bibfnamefont{A.~K.} \bibnamefont{Geim}},
  \bibinfo{journal}{Nat. Phys.} \textbf{\bibinfo{volume}{2}},
  \bibinfo{pages}{620} (\bibinfo{year}{2006}).

\bibitem[{\citenamefont{Mucciolo et~al.}(2009)\citenamefont{Mucciolo,
  Castro~Neto, and Lewenkopf}}]{Mucciolo2009}
\bibinfo{author}{\bibfnamefont{E.~R.} \bibnamefont{Mucciolo}},
  \bibinfo{author}{\bibfnamefont{A.~H.} \bibnamefont{Castro~Neto}},
  \bibnamefont{and} \bibinfo{author}{\bibfnamefont{C.~H.}
  \bibnamefont{Lewenkopf}}, \bibinfo{journal}{Phys. Rev. B}
  \textbf{\bibinfo{volume}{79}}, \bibinfo{pages}{075407}
  (\bibinfo{year}{2009}).

\bibitem[{\citenamefont{Schnez et~al.}(2009)\citenamefont{Schnez, Molitor,
  Stampfer, Güttinger, Shurobalko, Ihn, and Ensslin}}]{Schnez2009}
\bibinfo{author}{\bibfnamefont{S.}~\bibnamefont{Schnez}},
  \bibinfo{author}{\bibfnamefont{F.}~\bibnamefont{Molitor}},
  \bibinfo{author}{\bibfnamefont{C.}~\bibnamefont{Stampfer}},
  \bibinfo{author}{\bibfnamefont{J.}~\bibnamefont{Güttinger}},
  \bibinfo{author}{\bibfnamefont{I.}~\bibnamefont{Shurobalko}},
  \bibinfo{author}{\bibfnamefont{T.}~\bibnamefont{Ihn}}, \bibnamefont{and}
  \bibinfo{author}{\bibfnamefont{K.}~\bibnamefont{Ensslin}},
  \bibinfo{journal}{Appl. Phys. Lett.} \textbf{\bibinfo{volume}{94}},
  \bibinfo{pages}{012107} (\bibinfo{year}{2009}).

\bibitem[{\citenamefont{Stampfer et~al.}(2009)\citenamefont{Stampfer,
  Güttinger, Hellmüller, Molitor, Ensslin, and Ihn}}]{Stampfer2009}
\bibinfo{author}{\bibfnamefont{C.}~\bibnamefont{Stampfer}},
  \bibinfo{author}{\bibfnamefont{J.}~\bibnamefont{Güttinger}},
  \bibinfo{author}{\bibfnamefont{S.}~\bibnamefont{Hellmüller}},
  \bibinfo{author}{\bibfnamefont{F.}~\bibnamefont{Molitor}},
  \bibinfo{author}{\bibfnamefont{K.}~\bibnamefont{Ensslin}}, \bibnamefont{and}
  \bibinfo{author}{\bibfnamefont{T.}~\bibnamefont{Ihn}},
  \bibinfo{journal}{Phys. Rev. Lett.} \textbf{\bibinfo{volume}{102}},
  \bibinfo{pages}{056403} (\bibinfo{year}{2009}).

\bibitem[{\citenamefont{Güttinger et~al.}(2009)\citenamefont{Güttinger,
  Stampfer, Libisch, Frey, Burgdörfer, Ihn, and Ensslin}}]{Guettinger2009}
\bibinfo{author}{\bibfnamefont{J.}~\bibnamefont{Güttinger}},
  \bibinfo{author}{\bibfnamefont{C.}~\bibnamefont{Stampfer}},
  \bibinfo{author}{\bibfnamefont{F.}~\bibnamefont{Libisch}},
  \bibinfo{author}{\bibfnamefont{T.}~\bibnamefont{Frey}},
  \bibinfo{author}{\bibfnamefont{J.}~\bibnamefont{Burgdörfer}},
  \bibinfo{author}{\bibfnamefont{T.}~\bibnamefont{Ihn}}, \bibnamefont{and}
  \bibinfo{author}{\bibfnamefont{K.}~\bibnamefont{Ensslin}},
  \bibinfo{journal}{Phys. Rev. Lett.} \textbf{\bibinfo{volume}{103}},
  \bibinfo{pages}{046810} (\bibinfo{year}{2009}).

\bibitem[{\citenamefont{Neubeck et~al.}(2010)\citenamefont{Neubeck,
  Ponomarenko, Freitag, Giesbers, Zeitler, Morozov, Blake, Geim, and
  Novoselov}}]{Neubeck2010}
\bibinfo{author}{\bibfnamefont{S.}~\bibnamefont{Neubeck}},
  \bibinfo{author}{\bibfnamefont{L.~A.} \bibnamefont{Ponomarenko}},
  \bibinfo{author}{\bibfnamefont{F.}~\bibnamefont{Freitag}},
  \bibinfo{author}{\bibfnamefont{A.~J.~M.} \bibnamefont{Giesbers}},
  \bibinfo{author}{\bibfnamefont{U.}~\bibnamefont{Zeitler}},
  \bibinfo{author}{\bibfnamefont{S.~V.} \bibnamefont{Morozov}},
  \bibinfo{author}{\bibfnamefont{P.}~\bibnamefont{Blake}},
  \bibinfo{author}{\bibfnamefont{A.~K.} \bibnamefont{Geim}}, \bibnamefont{and}
  \bibinfo{author}{\bibfnamefont{K.~S.} \bibnamefont{Novoselov}},
  \bibinfo{journal}{Small} \textbf{\bibinfo{volume}{6}}, \bibinfo{pages}{1469}
  (\bibinfo{year}{2010}).

\bibitem[{\citenamefont{Wang et~al.}(2011{\natexlab{a}})\citenamefont{Wang,
  Cao, Tu, Li, Zhou, Hao, Guo, and Guo}}]{Wang2011a}
\bibinfo{author}{\bibfnamefont{L.-J.} \bibnamefont{Wang}},
  \bibinfo{author}{\bibfnamefont{G.}~\bibnamefont{Cao}},
  \bibinfo{author}{\bibfnamefont{T.}~\bibnamefont{Tu}},
  \bibinfo{author}{\bibfnamefont{H.-O.} \bibnamefont{Li}},
  \bibinfo{author}{\bibfnamefont{C.}~\bibnamefont{Zhou}},
  \bibinfo{author}{\bibfnamefont{X.-J.} \bibnamefont{Hao}},
  \bibinfo{author}{\bibfnamefont{G.-C.} \bibnamefont{Guo}}, \bibnamefont{and}
  \bibinfo{author}{\bibfnamefont{G.-P.} \bibnamefont{Guo}},
  \bibinfo{journal}{Chin. Phys. Lett.} \textbf{\bibinfo{volume}{28}},
  \bibinfo{pages}{067301} (\bibinfo{year}{2011}{\natexlab{a}}).

\bibitem[{\citenamefont{Wang et~al.}(2011{\natexlab{b}})\citenamefont{Wang,
  Ouyang, Jiao, Wang, Xie, Wu, Guo, and Dai}}]{Wang2011}
\bibinfo{author}{\bibfnamefont{X.}~\bibnamefont{Wang}},
  \bibinfo{author}{\bibfnamefont{Y.}~\bibnamefont{Ouyang}},
  \bibinfo{author}{\bibfnamefont{L.}~\bibnamefont{Jiao}},
  \bibinfo{author}{\bibfnamefont{H.}~\bibnamefont{Wang}},
  \bibinfo{author}{\bibfnamefont{L.}~\bibnamefont{Xie}},
  \bibinfo{author}{\bibfnamefont{J.}~\bibnamefont{Wu}},
  \bibinfo{author}{\bibfnamefont{J.}~\bibnamefont{Guo}}, \bibnamefont{and}
  \bibinfo{author}{\bibfnamefont{H.}~\bibnamefont{Dai}},
  \bibinfo{journal}{Nature Nanotech.} \textbf{\bibinfo{volume}{6}},
  \bibinfo{pages}{563} (\bibinfo{year}{2011}{\natexlab{b}}).

\bibitem[{Pla()}]{PlateCap}
\bibinfo{note}{We extract the backgate capacitance using $C_{BG} = e/\delta
  V_{BG} = e/(\alpha_{rel}\cdot\delta V_{SG})$ and compare it to a parallel
  plate capacitor with $C_{pp} = \epsilon \epsilon_{0}\cdot A / d$ where
  $\delta V_{BG}, \delta V_{SG}$ are the average voltages for adding one
  electron with the respective gate, $\alpha_{rel}$ is the relative side/back
  gate lever arm (from Fig.\,3a,b), $\epsilon=3.9$ for $SiO_{2}$ with thickness
  $d = 294\,$nm, $\epsilon_{0}$ is the vacuum permittivity, and $A$ is the dot
  area. We obtain $A=87^2\,$nm$^2$ and $A=59^2\,$nm$^2$, compared to AFM scan
  dimensions of $A=80\times90\,$nm$^2$ and $A=60\times67\,$nm$^2$,
  respectively, giving good agreement within the error bars of $\sim 15\%$}.

\bibitem[{\citenamefont{Dröscher et~al.}(2011)\citenamefont{Dröscher, Knowles,
  Meir, Ensslin, and Ihn}}]{Droescher2011}
\bibinfo{author}{\bibfnamefont{S.}~\bibnamefont{Dröscher}},
  \bibinfo{author}{\bibfnamefont{H.}~\bibnamefont{Knowles}},
  \bibinfo{author}{\bibfnamefont{Y.}~\bibnamefont{Meir}},
  \bibinfo{author}{\bibfnamefont{K.}~\bibnamefont{Ensslin}}, \bibnamefont{and}
  \bibinfo{author}{\bibfnamefont{T.}~\bibnamefont{Ihn}},
  \bibinfo{journal}{Phys. Rev. B} \textbf{\bibinfo{volume}{84}},
  \bibinfo{pages}{073405} (\bibinfo{year}{2011}).

\bibitem[{\citenamefont{Molitor
  et~al.}(2009{\natexlab{b}})\citenamefont{Molitor, Dröscher, Güttinger,
  Jacobsen, Stampfer, Ihn, and Ensslin}}]{Molitor2009a}
\bibinfo{author}{\bibfnamefont{F.}~\bibnamefont{Molitor}},
  \bibinfo{author}{\bibfnamefont{S.}~\bibnamefont{Dröscher}},
  \bibinfo{author}{\bibfnamefont{J.}~\bibnamefont{Güttinger}},
  \bibinfo{author}{\bibfnamefont{A.}~\bibnamefont{Jacobsen}},
  \bibinfo{author}{\bibfnamefont{C.}~\bibnamefont{Stampfer}},
  \bibinfo{author}{\bibfnamefont{T.}~\bibnamefont{Ihn}}, \bibnamefont{and}
  \bibinfo{author}{\bibfnamefont{K.}~\bibnamefont{Ensslin}},
  \bibinfo{journal}{Appl. Phys. Lett.} \textbf{\bibinfo{volume}{94}},
  \bibinfo{pages}{222107} (\bibinfo{year}{2009}{\natexlab{b}}).

\bibitem[{\citenamefont{van~der Viel et~al.}(2003)\citenamefont{van~der Viel,
  De~Franceschi, Elzerman, Fujisawa, Tarucha, and Kouwenhoven}}]{DDreview}
\bibinfo{author}{\bibfnamefont{W.~G.} \bibnamefont{van~der Viel}},
  \bibinfo{author}{\bibfnamefont{S.}~\bibnamefont{De~Franceschi}},
  \bibinfo{author}{\bibfnamefont{J.~M.} \bibnamefont{Elzerman}},
  \bibinfo{author}{\bibfnamefont{T.}~\bibnamefont{Fujisawa}},
  \bibinfo{author}{\bibfnamefont{S.}~\bibnamefont{Tarucha}}, \bibnamefont{and}
  \bibinfo{author}{\bibfnamefont{L.~P.} \bibnamefont{Kouwenhoven}},
  \bibinfo{journal}{Rev. Mod. Phys.} \textbf{\bibinfo{volume}{75}},
  \bibinfo{pages}{1} (\bibinfo{year}{2003}).

\bibitem[{\citenamefont{Recher et~al.}(2009)\citenamefont{Recher, Nilsson,
  Burkard, and Trauzettel}}]{Recher2009}
\bibinfo{author}{\bibfnamefont{P.}~\bibnamefont{Recher}},
  \bibinfo{author}{\bibfnamefont{J.}~\bibnamefont{Nilsson}},
  \bibinfo{author}{\bibfnamefont{G.}~\bibnamefont{Burkard}}, \bibnamefont{and}
  \bibinfo{author}{\bibfnamefont{B.}~\bibnamefont{Trauzettel}},
  \bibinfo{journal}{Phys. Rev. B.} \textbf{\bibinfo{volume}{79}},
  \bibinfo{pages}{085407} (\bibinfo{year}{2009}).

\bibitem[{\citenamefont{Libisch et~al.}(2010)\citenamefont{Libisch, Rotter,
  Güttinger, Stampfer, and Burgdörfer}}]{Libisch2010}
\bibinfo{author}{\bibfnamefont{F.}~\bibnamefont{Libisch}},
  \bibinfo{author}{\bibfnamefont{S.}~\bibnamefont{Rotter}},
  \bibinfo{author}{\bibfnamefont{J.}~\bibnamefont{Güttinger}},
  \bibinfo{author}{\bibfnamefont{C.}~\bibnamefont{Stampfer}}, \bibnamefont{and}
  \bibinfo{author}{\bibfnamefont{J.}~\bibnamefont{Burgdörfer}},
  \bibinfo{journal}{Phys. Rev. B.} \textbf{\bibinfo{volume}{81}},
  \bibinfo{pages}{245411} (\bibinfo{year}{2010}).

\bibitem[{\citenamefont{Folk et~al.}(2001)\citenamefont{Folk, Marcus,
  Berkovits, Kurland, Aleiner, and Altshuler}}]{Folk2001}
\bibinfo{author}{\bibfnamefont{J.~A.} \bibnamefont{Folk}},
  \bibinfo{author}{\bibfnamefont{C.~M.} \bibnamefont{Marcus}},
  \bibinfo{author}{\bibfnamefont{R.}~\bibnamefont{Berkovits}},
  \bibinfo{author}{\bibfnamefont{I.~L.} \bibnamefont{Kurland}},
  \bibinfo{author}{\bibfnamefont{I.~L.} \bibnamefont{Aleiner}},
  \bibnamefont{and} \bibinfo{author}{\bibfnamefont{B.~L.}
  \bibnamefont{Altshuler}}, \bibinfo{journal}{Phys. Scr.}
  \textbf{\bibinfo{volume}{T90}}, \bibinfo{pages}{26} (\bibinfo{year}{2001}).

\bibitem[{\citenamefont{Lundeberg and Folk}(2010)}]{Lundeberg2010}
\bibinfo{author}{\bibfnamefont{M.~B.} \bibnamefont{Lundeberg}}
  \bibnamefont{and} \bibinfo{author}{\bibfnamefont{J.~A.} \bibnamefont{Folk}},
  \bibinfo{journal}{Phys. Rev. Lett.} \textbf{\bibinfo{volume}{105}},
  \bibinfo{pages}{146804} (\bibinfo{year}{2010}).

\bibitem[{\citenamefont{Lundeberg and Folk}(2009)}]{Lundeberg2009}
\bibinfo{author}{\bibfnamefont{M.~B.} \bibnamefont{Lundeberg}}
  \bibnamefont{and} \bibinfo{author}{\bibfnamefont{J.~A.} \bibnamefont{Folk}},
  \bibinfo{journal}{Nat. Phys.} \textbf{\bibinfo{volume}{5}},
  \bibinfo{pages}{894} (\bibinfo{year}{2009}).

\bibitem[{\citenamefont{Rao et~al.}(2011)\citenamefont{Rao, Stesmans, Keunen,
  Kosynkin, Higginbotham, and Tour}}]{Rao2011}
\bibinfo{author}{\bibfnamefont{S.~S.} \bibnamefont{Rao}},
  \bibinfo{author}{\bibfnamefont{A.}~\bibnamefont{Stesmans}},
  \bibinfo{author}{\bibfnamefont{K.}~\bibnamefont{Keunen}},
  \bibinfo{author}{\bibfnamefont{D.~V.} \bibnamefont{Kosynkin}},
  \bibinfo{author}{\bibfnamefont{A.}~\bibnamefont{Higginbotham}},
  \bibnamefont{and} \bibinfo{author}{\bibfnamefont{J.~M.} \bibnamefont{Tour}},
  \bibinfo{journal}{Appl. Phys. Lett.} \textbf{\bibinfo{volume}{98}},
  \bibinfo{pages}{083116} (\bibinfo{year}{2011}).

\bibitem[{\citenamefont{Tombros et~al.}(2011)\citenamefont{Tombros, Veligura,
  Junesch, Guimaraes, Vera-Marun, Jonkman, and van Wees}}]{Tombros2011}
\bibinfo{author}{\bibfnamefont{N.}~\bibnamefont{Tombros}},
  \bibinfo{author}{\bibfnamefont{A.}~\bibnamefont{Veligura}},
  \bibinfo{author}{\bibfnamefont{J.}~\bibnamefont{Junesch}},
  \bibinfo{author}{\bibfnamefont{M.~H.~D.} \bibnamefont{Guimaraes}},
  \bibinfo{author}{\bibfnamefont{I.~J.} \bibnamefont{Vera-Marun}},
  \bibinfo{author}{\bibfnamefont{H.~T.} \bibnamefont{Jonkman}},
  \bibnamefont{and} \bibinfo{author}{\bibfnamefont{B.~J.} \bibnamefont{van
  Wees}}, \bibinfo{journal}{Nature Phys.} \textbf{\bibinfo{volume}{7}},
  \bibinfo{pages}{697} (\bibinfo{year}{2011}).

\bibitem[{\citenamefont{Moreno-Moreno et~al.}(2009)\citenamefont{Moreno-Moreno,
  Castellanos-Gomez, Rubio-Bollinger, Gomez-Herrero, and Agrait}}]{Moreno2009}
\bibinfo{author}{\bibfnamefont{M.}~\bibnamefont{Moreno-Moreno}},
  \bibinfo{author}{\bibfnamefont{A.}~\bibnamefont{Castellanos-Gomez}},
  \bibinfo{author}{\bibfnamefont{G.}~\bibnamefont{Rubio-Bollinger}},
  \bibinfo{author}{\bibfnamefont{J.}~\bibnamefont{Gomez-Herrero}},
  \bibnamefont{and} \bibinfo{author}{\bibfnamefont{N.}~\bibnamefont{Agrait}},
  \bibinfo{journal}{Small} \textbf{\bibinfo{volume}{5}}, \bibinfo{pages}{924}
  (\bibinfo{year}{2009}).

\bibitem[{\citenamefont{Shi et~al.}(2011)\citenamefont{Shi, Yang, Zhang, Wang,
  Liu, Shi, Wang, and Zhang}}]{Shi2011}
\bibinfo{author}{\bibfnamefont{Z.}~\bibnamefont{Shi}},
  \bibinfo{author}{\bibfnamefont{R.}~\bibnamefont{Yang}},
  \bibinfo{author}{\bibfnamefont{L.}~\bibnamefont{Zhang}},
  \bibinfo{author}{\bibfnamefont{Y.}~\bibnamefont{Wang}},
  \bibinfo{author}{\bibfnamefont{D.}~\bibnamefont{Liu}},
  \bibinfo{author}{\bibfnamefont{D.}~\bibnamefont{Shi}},
  \bibinfo{author}{\bibfnamefont{E.}~\bibnamefont{Wang}}, \bibnamefont{and}
  \bibinfo{author}{\bibfnamefont{G.}~\bibnamefont{Zhang}},
  \bibinfo{journal}{Adv. Mater.} \textbf{\bibinfo{volume}{23}},
  \bibinfo{pages}{3061} (\bibinfo{year}{2011}).

\end{thebibliography}

\end{document}